\documentclass[12pt,a4]{article}

\usepackage{amsthm,amsmath,latexsym,amssymb,amsfonts,amscd}
\usepackage{graphics,lscape,fancyhdr,array,stmaryrd,euscript}
\usepackage{empheq}
\usepackage{dsfont}
\usepackage{verbatim}
\usepackage{color,tikz,tikz-cd}
\usetikzlibrary[snakes]
\usepackage{relsize,slashed}
\numberwithin{equation}{section}
\usepackage{hyperref}

\newcommand{\bep}{\begin{picture}}
\newcommand{\eep}{\end{picture}}

\newcounter{YoungHeight}\newcounter{YoungWidth}

\newcounter{Mul1}\newcounter{Mul2}\newcounter{Mul3}\newcounter{Mul4}
\newcounter{A0}\newcounter{A1}\newcounter{A2}
\newcounter{B3}
\newcounter{C3}\newcounter{C4}
\newcounter{D1}\newcounter{D2}\newcounter{D3}
\newcounter{T0}\newcounter{T1}

\newlength{\txtHShift}

\newlength{\txtWidth}

\newcommand{\Add}[3]{\setcounter{#1}{#2}\addtocounter{#1}{#3}}

\newcommand{\Length}[1]{#10}
\newcommand{\YoungScale}{}

\newcommand{\BlockApar}[2]{\parbox{\Length{#1}pt}{\YoungScale\bep(\Length{#1},\Length{#2}){\Add{A1}{#1}{1}\Add{A2}{#2}{1}}%
\multiput(0,0)(10,0){\value{A1}}{\line(0,1){\Length{#2}}}\multiput(0,0)(0,10){\value{A2}}{\line(1,0){\Length{#1}}}%
\setcounter{YoungHeight}{\Length{#2}}\setcounter{YoungWidth}{\Length{#1}}\eep}}

\newcommand{\YoungpA}{\BlockApar{1}{1}}
\newcommand{\YoungpB}{\BlockApar{2}{1}}

\newcommand{\YoungpAA}{\BlockApar{1}{2}}

\usepackage{setspace}

 \usepackage[numbers,sort&compress]{natbib}
 \usepackage[nottoc]{tocbibind}

\newcommand{\besubeqs}{\begin{subequations}}
\newcommand{\esubeqs}{\end{subequations}}

\newcommand{\mm}{{\ensuremath{{\mu}}}}

\newcommand{\fud}[2]{{}^{#1}{}_{#2}\,}
\newcommand{\fdu}[2]{{}_{#1}{}^{#2}\,}

\newcommand{\Tr}{{\mathrm{Tr}\,}}

\newcommand{\ttt}{{\boldsymbol{t}}}

\newcommand{\hs}{\mathfrak{hs}}

\begin{document}
\pagenumbering{gobble}
\hfill
\vspace{-1.5cm}
\vskip 0.05\textheight
\begin{center}
{\Large\bfseries Conformal Carrollian Spin-3 Gravity in 3d
}

\vspace{0.4cm}

\vskip 0.03\textheight
Iva \textsc{Lovrekovic}${}$

\vskip 0.03\textheight

\vspace{5pt}

{\it Institute for Theoretical Physics, Technische Universit\"at Wien, \\ Wiedner Hauptstrasse, 8-10,\\
1040, Wien, Austria}

\end{center}

\vskip 0.02\textheight

\begin{abstract}
We consider ultra-relativistic limit of the spin-2 and spin-3 conformal gravity theories in three dimensions, which leads to conformal Carrollian spin-2 gravity and its spin-3 analog.  We also comment on the generalization of the result to arbitrary spin. The holographic examples show generalization in comparison with non-conformal Carrollian gravity.
\end{abstract}

\newpage
\pagenumbering{arabic}
\setcounter{page}{2}

\section{Introduction}

The motivation for the research of conformal Carrollian Spin-3 Gravity in 3d comes from several sides. Conformal symmetry has been studied from  number of aspects, as a symmetry that brings extra simplifications, defines gravity models as a toy models using which one can resolve otherwise unresolvable issues, or as a symmetry that plays a key role in the Physics at the Planck's energy scale \cite{tHooft:2014swy}.

Carroll symmetries have been studied due to the connection with asymptotic symmetries of flat-space times described by Bondi-Metzner-Sachs (BMS) group and their relation with near horizon boundary conditions \cite{Afshar:2016wfy,Afshar:2016kjj}.  BMS symmetries are isomorphic to conformal Carrollian symmetries \cite{Bagchi:2010zz,Duval:2014uva}, and there is also a very interesting picture which relates soft theorems and asymptotic symmetries in QFT in asymptotically flat space-times with memory effects. Review of advances in this field can be found in \cite{Strominger:2017zoo}. Carroll algebra can be obtained as an ultra-relativistic limit of Poincare algebra, and similarly conformal version follows from conformal algebra  \cite{Bagchi:2019xfx}. The $c\to 0$ limit is depicted in a light cone structure which has a light cone collapsed into a line \cite{Hartong:2015xda}. The Carrollian space-times are formed in future and past null-infinity and conformal Carroll algebra can be viewed as conformal extension of BMS algebra. 
  In the near horizon boundary conditions one obtains BMS algebra as a composite in the terms of infinite copies of Heisenberg algebra. 

Carroll symmetries have recently been studied in relation with higher spin theories \cite{Campoleoni:2021blr} with the aim of filling the gap  of the less developed field of the higher spin extensions of the Poincare Algebra in higher than three dimensions. In three dimensions the situation is better, where number of constructions has been considered  \cite{Blencowe:1988gj,Campoleoni:2011tn,Afshar:2013vka}.
At present, there is a handful of complete higher spin gravity models that have actions and avoid the non-locality issue: (a) $3d$ massless \cite{Blencowe:1988gj,Campoleoni:2010zq,Henneaux:2010xg}; (b) $4d$ conformal \cite{Segal:2002gd,Tseytlin:2002gz}; (c) $4d$ chiral \cite{Ponomarev:2016lrm} and (d) $3d$ conformal \cite{Pope:1989vj}.

Higher spin gravities in three dimensions can be constructed consistently while containing a finite number of higher spin-fields. In fact, one can construct infinite number of such theories \cite{Grigoriev:2019xmp}.
In the relativistic theory, this is allowed by considering the action of Chern-Simons form which has also been used in construction of the non-conformal Carrollian theories \cite{Bergshoeff:2016soe}. Since higher spin theories usually contain infinite number of higher spin fields and only several consistent theories are known, it is our aim to use one of them \cite{Grigoriev:2019xmp} and investigate what can the ultra-relativistic limit teach us.

In the construction of the gauge theories based on the Carrollian algebras, 
the ultra-relativistc limit of General relativity in the first order formulation gives equations of motion which are not sufficient to find all the components of the connection fields expressed in terms of the other fields. Those components play a role of Lagrange multipliers for the constraints on the geometry \cite{Bergshoeff:2017btm}, and carry a physical interpretation in terms of radiative degrees of freedom for gravity at null-distances \cite{Herfray:2021qmp,Herfray:2020rvq}. 
Which is what we can expect to appear here as well.

The paper is structured as follows: In the first section we review the construction using which we obtain spin-2 and spin-3 cases of conformal gravities, we review conformal gravity in three dimensions, i.e. linearization of the theory and the ultra-relativistic limit. In the third section we analyze the ultra-relativistic limit of the spin-2 case while in the fourth section we analyze the spin-3 case and comment generalization to arbitrary spin. In the fifth section we consider the holography of the theories, and in the sixth section we conclude. 

\section{Construction}

We study Chern-Simons action  
\begin{align}
    S[\omega]&=\int \Tr\left[\omega \wedge d \omega +\frac23 \omega\wedge \omega\wedge \omega\right]\,. \label{cs}
\end{align}
as a gauge theory for the Carollian conformal algebra.
Knowing the higher spin extension of $so(3,2)$, (i.e. algebras that have $so(3,2)$ as a subalgebra and also contain  other nontrivial representations of $so(3,2)$), we can take the ultra-relativistic limit that gives conformal  Carrollian algebra and consider their generalization to the higher spin. 

To construct finite spectrum of higher spin fields (called Fradkin-Tseytlin fields \cite{Fradkin:1985am}) one needs to take a  nontrivial  finite-dimensional irreducible representation $V$ of $so(3,2)$, such as an irreducible tensor or a spin-tensor. Then one needs to evaluate $U(so(d,2))$ in $V$, which means multiply the generators \footnote{Here, indices $A,B,...=0,...,4$ denote the indices of $so(3,2)$ while  $\eta_{AB}$ denotes invariant metric.} $T_{AB}=-T_{BA}$ of $so(3,2)$ in given representation and determine the generated algebra  \cite{Grigoriev:2019xmp}.
In general, this algebra is denoted with $\hs(V)$. Here we are going to consider spin-2 and spin-3 fields. 
For that case, one needs to take the vector representation, which is denoted by one-cell Young diagram $\YoungpA$. Then, the spectrum of the corresponding algebra, $\hs(\YoungpA)$ is  the following:
\begin{align}
    \mathfrak{hs}(\YoungpA)&= \YoungpA\otimes \YoungpA= \YoungpAA \oplus \YoungpB\oplus \bullet .
\end{align}
This algebra is a matrix algebra with $(3+2)^2$ generators $t\fdu{A}{B}$ which are decomposed with respect to $so(3,2)$. 
The $gl_{3+2}$ commutation relations 
\begin{align}
    [t\fdu{A}{B}, t\fdu{C}{D}]&=-\delta\fdu{A}{D}t\fdu{C}{B} +\delta\fdu{C}{B} t\fdu{A}{D}\,. \label{alg2}
\end{align}
with the irreducible generators  $T_{AB}=-T_{BA}$, $S_{AB}=S_{BA}$ and $R$,
 in the $so(3,2)$ base
\begin{align}
    T_{AB}&=t_{A|B}-t_{B|A}\,, & S_{AB}&=t_{A|B}+t_{B|A}-\frac{2}{d+2}t\fdu{C}{C}\eta_{AB}\,, &&R=t\fdu{C}{C}\,, \label{eq15}
\end{align}
lead to  the commutation relations
\besubeqs
\begin{align}
    [T_{AB},T_{CD}]&= \eta_{BC} T_{AD}-\eta_{AC} T_{BD}-\eta_{BD} T_{AC}+\eta_{AD} T_{BC}\,,\\
    [T_{AB},S_{CD}]&= \eta_{BC}S_{AD}-\eta_{AC}S_{BD}+\eta_{BD}S_{AC}-\eta_{AD}S_{BC}\,,\\
    [S_{AB},S_{CD}]&= \eta_{BC}T_{AD}+\eta_{AC}T_{BD}+\eta_{BD}T_{AC}+\eta_{AD}T_{BC}\,.
\end{align}
\esubeqs
Here $R$ commutes with everything because it is associated with $1$ in $gl(V)$. So it can be truncated away. One obtains a  theory that contains two fields:
\begin{align}
    \omega&= \omega^{A,B} T_{AB} +\omega^{AB} S_{AB}
\end{align}
$T_{AB}$ is the conformal graviton and $S_{AB} $ is a field similar to the spin-three partially-massless field. From the bilinear form of $gl_n$
\begin{align}
    \Tr (t\fdu{A}{B}, t\fdu{C}{D})&=\delta\fdu{A}{D}\delta\fdu{C}{B}
\end{align}
one obtains invariant bilinear forms for $so(3,2)$
\begin{align}
    \Tr (T_{AB}, T_{CD})&=2(\eta_{AD}\eta_{CB}-\eta_{BD}\eta_{CA})\\
    \Tr (S_{AB}, S_{CD})&=2(\eta_{AD}\eta_{CB}+\eta_{BD}\eta_{CA}-\frac{2}{d+2}\eta_{AB}\eta_{CD})
\end{align}

Before proceeding to linearizing a theory and taking the ultra-relativistic limit,
let us first remember conformal gravity as a gauge theory in three dimensions, studied in \cite{Horne:1988jf}. 
Fixing the commutation relations of the $so(3,2)$ conformal algebra (for $a=0,1,2$)
    \besubeqs\label{LPDK}
\begin{align}
[D,P^a]&=-P^a\,, & [J^{ab},P^c]&=P^a\eta^{bc}-P^b\eta^{ac}\,,\\
[D,K^a]&=K^a\,,  & [J^{ab},K^c]&=K^a\eta^{bc}-K^b\eta^{ac}\,,\\
[P^a,K^b]&=-J^{ab}+\eta^{ab}D\,, &[J^{ab},J^{cd}]&=J^{ad}\eta^{bc}-J^{ac}\eta^{bd}-J^{bd}\eta^{ac}+J^{bc}\eta^{ad},,\label{confalg}
\end{align}
\esubeqs
for $P^a$ translations, $K^a$ special conformal tranformations, $J^{ab}$ Lorentz boosts and $D$ dilatation, we can write the connection for the $so(3,2)$ as
\begin{align}
    \omega&= \frac12 \varpi^{a,b} J_{ab}+e^aP_a +f^a K_a+ bD.
\end{align}
The action, is a standard Chern-Simons action
\begin{align}
    S[\omega]&=\int \Tr\left[\omega \wedge d \omega +\frac{2}{3} \omega\wedge \omega\wedge \omega\right]\,\label{csac3d}
\end{align}
where the curvature $F=d\omega+\frac12[\omega,\omega]$ is set to zero by the equations.
The components of the curvature give
\besubeqs
\begin{align}
    F[P^{a}]&=\nabla e^a- b\wedge e^a,\\
    F[D]&= \nabla b+ e_m \wedge f^m,\\
    F[J^{a,b}]&= R^{a,b}-e^a\wedge f^b+e^b\wedge f^a,\\
    F[K^a]&=\nabla f^a+b\wedge f^a,
\end{align}
\esubeqs
where we denote the Lorentz covariant derivative with $\nabla =d+\varpi$. In this notation Riemann two-form is given by
 $R^{a,b}=d\omega^{a,b}+\omega\fud{a,}{c}\wedge \omega^{c,b}$. To show how one obtains the theory when $\omega$ from the action (\ref{csac3d}) is only a spin connection and $e_{\mu}{}^a$ only dynamical variable, we have to solve the equations and fix the gauge conditions.
Our gauge parameter is a $so(3,2)$ algebra valued 0-form, 
\begin{align}
    \Xi&= \frac12 \eta^{a,b} L_{ab}+\xi^aP_a +\zeta^a K_a+ \rho D
\end{align}
which gives
\besubeqs
\begin{align}
    \delta e^a&=\nabla \xi^a - \rho\wedge e^a-b\wedge \xi^a+\eta^{a,}{}_b \wedge e^a ,\\\
    \delta b&= \nabla \rho +\xi_m \wedge f^m+e_m \wedge \zeta^m,\\
    \delta\omega^{a,b}&= \nabla \eta^{a,b}-\xi^a\wedge f^b+\xi^b\wedge f^a-e^a\wedge \zeta^b+e^b\wedge \zeta^a,\\
    \delta f^a&=\nabla  \zeta^a+b\wedge \zeta^a+\rho\wedge f^a+ \eta^{a,}{}_b \wedge f^a\,.
\end{align}
\esubeqs
If we assume that the dreibein $e^a_\mm$ is invertible, we can fix the gauge to set $b=0$. Now, one can write  $g_{\mu\nu}=e^a_\mu e^b_\nu \eta_{ab}$ for the conformal metric. From the equation $F[D]=0$ we obtain that $f^a_\mu e_{a\nu}$ is symmetric. The equation $F[P^a]=0$ we recognize as the standard torsion constraint, which defines  $\varpi^{a,b}$ using $e^a$. The following equation $F[J^{a,b}]=0$ defines that $f^a_\mu e_{a\nu}$ is the Schouten tensor, while the $F[K^a]=0$ sets Cotton tensor $C_{\mu\nu}=0$. That equation is the only equation which is dynamical. 
If we plug these solutions for the fields back into the action \eqref{csac3d} we obtain the action of the same form, where the only dynamical variable is $e^a$. 

Here, we want to determine the fields of this conformal gravity theory for the conformal graviton and a spin three field in an ultrarelativistic limit, i.e. when we have conformal Carrollian theory. To find the spectrum of the fields, one has to linearize a theory over the Minkowski vacuum. We choose it to be $\omega_0=h^aP_a$ for $h^a=h_{\mu}{}^adx^{\mu}$, where the background metric is $h_{\mu}{}^a=\delta_{\mu}{}^a$. The equations at the linear order and the linearized gauge symmetries are 
\begin{align}
    d\omega+\omega_0\wedge\omega+\omega\wedge\omega_0=0, && \delta\omega =d\xi+[\omega_0,\xi].\label{lineq}
\end{align}
Here, $\omega_0$ is the background field, while $\omega$ field is the Lie algebra valued one-form. It is valued in one of the algebras of the so(3,2) irreducible modules obtained from the decomposition of the $\mathfrak{hs}(V)$. In our case it will be valued in the ultra-relativistic limit of the algebra we consider: in the ultra-relativistic limit of the conformal algebra and in the ultra-relativistic limit of the extension of the conformal algebra for spin-three field described above.

For the generators $t^{\Lambda}$ the equations (\ref{lineq}), with $\omega=\omega^{\Lambda}t_{\Lambda}$ and $\xi=\xi^{\Lambda}t_{\Lambda}$, will give 
\begin{align}
    d\omega^{\Lambda}t_{\Lambda}+h^a\wedge\omega^{\Lambda}[P_a,t_{\Lambda}]=0 && \delta \omega^{\Lambda}t_{\Lambda}=d\xi^{\Lambda}t_{\Lambda}+h^a\wedge\xi^{\Lambda}[P_a,t_{\Lambda}]
\end{align}
where $t^{\Lambda}$ denotes the generators of the so(3,2) irreducible modules $T_{AB}$, $S_{AB}$ etc.,  $\Lambda$ denotes indices in the light cone coordinates $\Lambda=a,+,-$, and $\eta^{AB}$ is corresponding metric with $\eta_{+-}=\eta_{-+}=1$. 

To take the ultra-relativistic limit we are going to follow the procedure of taking In\"on\"u-Wigner (IW) contraction \cite{Inonu:1953sp}.
In our terminology we do this as follows.
We denote the Lie algebra from which we start with $\mathfrak{g}$, while $\mathfrak{h}$ is a subalgebra. The decomposition $\mathfrak{g}=\mathfrak{h}+\mathfrak{i}$ denotes a direct sum of vector spaces. The generators of the ideal $\mathfrak{i}$, are rescaled with contraction parameter $\epsilon$, such that $\mathfrak{i}\rightarrow\epsilon\mathfrak{i}$. The commutation relations then take the form
\begin{align}
    [\mathfrak{h},\mathfrak{h}]=\mathfrak{h}\,,  &&  [\mathfrak{h},\mathfrak{i}]=\frac{1}{\epsilon}\mathfrak{h}+\mathfrak{i}\,, && [\mathfrak{i},\mathfrak{i}]=\frac{1}{\epsilon^2}\mathfrak{h}+\frac{1}{\epsilon}\mathfrak{i}
\end{align}
which means that when taking
$\epsilon\to\infty$, one obtains well defined limit. For this to hold, $\mathfrak{h}$ needs to be subalgebra of $\mathfrak{g}$. If one had $[\mathfrak{h},\mathfrak{h}]=\mathfrak{h}+\epsilon \mathfrak{i}$ the limit when $\epsilon\to\infty$ would not be well defined.

\section{Spin-2}

Let us focus on the first case with a Young module $\YoungpA$ for conformal graviton when $t_{ab}=-t_{ab}$. The dictionary between the generators $T_{AB}$ for the conformal graviton and generators of translations, special conformal transformations, dilatation and Lorentz boosts is $P_a=T_{a+}$, $K_a=T_{a-}$, $D=-T_{+-}$, $L_{ab}=T_{ab}$ respectively.
To obtain the generators in the ultra-relativistic limit from the generators $t^{\Lambda}$, we perform an IW contraction \cite{Inonu:1953sp} of the underlying algebra, in a way that leads to the conformal Carrollian algebra.

We introduce space-time splitting of indices where 
$a=\{0,i; i=1,2\}$. That allows us to introduce the notation
\begin{align}
    P_0\equiv H\,, && K_0\equiv K_0\,, && J_{0i}\equiv B_i
\end{align}
When we choose the subalgebra $\mathfrak{h}$ and ideal $\mathfrak{i}$ as
\begin{align}
\mathfrak{h}=\{P_i,K_i,D,J_{ij}\}\,, && \mathfrak{i}=\{H,K_0,B_i\}
\end{align}
starting from the conformal algebra (\ref{confalg}), we obtain the algebra
\besubeqs
\begin{align}
[D,H]&=-H\,, & [D,K_0]&=K_0\,, \\
[B^{j},P^k]&=H\eta^{jk}\,, & [B^{j},K^k]&=K^0\eta^{jk}\\
[H,K^j]&=-B^{j}\,, & [P^i,K^0]&=B^{i}\,,\\
[J^{ij},B^{l}]&=B^{i}\eta^{jl}-B^{j}\eta^{il}, & & \label{carca}
\end{align}
\esubeqs
i.e., conformal Carrollian algebra \cite{Bagchi:2019xfx}. 
The Lie algebra valued one-form now becomes
\begin{align}
    \omega&= e^i P_i+\tau H +\tfrac12 \omega^{i,j}J_{ij}+\beta^{j}B_{j}-b D+f^{i}K_{i}+\kappa K_{0}\label{connection}
\end{align}
with an analogous decomposition of the gauge parameter
\begin{align}
    \xi&= e_i \xi^{i+}+\tau \xi^{0+} +\tfrac12 \omega_{i,j}\xi^{i,j}+\beta_{j}\xi^{0j}+b \xi^{+-}+f_{i}\xi^{i-}+\kappa \xi^{0-}.\label{gaugeparam}
\end{align}
\besubeqs
The linearized equations of the conformal Carrollian gravity and the gauge symmeteries read
\begin{align}
    D&: &db-h_m \wedge f^{m}&=0\,, &\delta \omega^{+-}&=d\xi^{+-}-h_m \xi^{m-}\,, \label{etpm}\\ 
    P_{i}&: &de^{i}+h_m \wedge \omega^{m,i}-h^i\wedge b&=0\,, &\delta e^{i}&=d\xi^{i+}+h_m \xi^{m,i}-h^{i}\xi^{+-}\,,\label{etip}\\
    H&: &d\tau-h_i \wedge \beta^{i}-h^0\wedge b&=0\,, &\delta \tau&=d\xi^{0+}+h_i \xi^{0,i}-h^{0}\xi^{+-}\,,\label{et0p}\\
    t_{ij}&: &d\omega^{i,j}-h^i \wedge f^{j}+h^j \wedge f^{i}&=0\,, &\delta \omega^{i,j}&=d\xi^{i,j}-h^i \xi^{j-}+h^j \xi^{i-}\,,\label{etij}\\ 
    B_{j}&: &d\beta^{j}+h^j \wedge \kappa -h^0 \wedge f^{j}&=0\,, &\delta \beta^{j}&=d\xi^{0,j}-h^0 \xi^{j-}+h^j \xi^{0-}\,,\label{et0j}\\ 
    K_i&: &df^{i}&= 0\,, &\delta f^{i}&=d\xi^{i-} \,.\label{etim}\\
    t_{0-}&: &d\kappa&= 0\,, &\delta \kappa&=d\xi^{0-} \,. \label{eti0}
\end{align}
\esubeqs
To find the spectrum of the theory one has to solve this system of equations. From (\ref{etpm})
the gauge invariance allows to fix $b_{m}$ just like in three dimensions. 
That defines $\xi^{-}_{m}=\partial_m\xi^{+-}$, and makes $b_{m}=0$. 
The equation (\ref{etpm})
 demands that $f_{mn}=f_{nm}$ is symmetric field. The components $m,0$ of equation (\ref{etpm})
$ E[t_{+-}]^{m0}= -  h^{ma} f{}^{0}{}_{a} + \partial^{m}b{}^{0} =0 $ give \begin{align}f^{0}{}^m=\partial^m b^0. \label{ef0m} \end{align}
The gauge transformations (\ref{etip}) allow to fix antisymmetric part of $e{}_{[m|n]}=0$ and its trace $e{}_{m}{}^m=0$ using $\xi_m,{}^n=-\partial_m\xi^{+n}+h_m{}^n\xi^{+-}$ and $\xi^{+-}=\frac{1}{2}\partial_m\xi^{+m}$ respectively, and the remaining symmetric part we call $e_{(m,n)}\equiv\phi_{mn}$.
The corresponding equation 
\begin{align}
    E[t_{i+}]^{mni}=- \omega^{mni} + \omega^{nmi} + \partial^{m}\phi^{ni} -  \partial^{n}\phi^{mi}
\end{align}
can be split into symmetric, antisymmetric and hook part. 
The only non-trivial part, hook part, of the equation defines \begin{align}
    \omega^{jm,n}&=\partial^m\phi^{nj}-\partial^{n}\phi^{mj}\,, & \omega^{jm,}{}_{j}=\partial_j\phi^{mj}  \label{wabc}
\end{align}
 while the antisymmetric and symmetric part are exactly equal to zero. 
 The $m,0$ component of  (\ref{etip})  does not allow for any more gauge fixing. 
 The symmetric part of the equation gives $b^{0}=\frac{1}{2}\partial_i e^{0i}$
while the antisymmetric part leads to 
\begin{align}
    \omega^{0im}=\tfrac{1}{2} (-  \partial^{i}e^{0}{}^{m} + \partial^{m}e^{0}{}^{i}). \label{ew0im}
\end{align}
From the gauge transformation (\ref{et0p}) one can fix $\tau_m{}=0$ with $\xi^{0,}{}_m=\partial_m\xi^{0+}$.
The $m,n$ component of the corresponding equation 
$
    E[t_{0+}]^{mn}=-\beta^{nm}+\beta^{mn}=0
$
makes the field $\beta^{m,n}$ symmetric, 
while the $m,0$ component defines $\partial^m\tau^0{}=\beta^{0}{}^m$.

The gauge (\ref{etij}) has all the components of the gauge parameters fixed, which means the equation is
\begin{align}
    E[t_{ij}]^{mn}=- h^{nj} f{}^{mi} + h^{ni} f{}^{mj} + h^{mj} f{}^{ni} -  h^{mi} f{}^{nj} + \partial^{m}\omega^{nij} -  \partial^{n}\omega^{mij}=0.\label{fc}
\end{align}
The symmetric part of the equation exactly vanishes and does not give any conditions or definitions of the field. The antisymmetric part of the equation as well vanishes when the definition of $\omega^{ni,j}$ is used. The Riemann component of the equation can be contracted with the $h_{ij}$ to define
\begin{align}
    f^{i}{}_i=-\partial_j\omega_i{}^{ij}.
\end{align}
The Riemann part itself gives equation  constraint on the geometry which we express in terms of the fields $\beta^{ij},\tau^0,\phi^{ij}$, in (\ref{eqap}) the appendix.

The $m,0$ components of the equation define 
\begin{align}
E[t_{ij}]^{m0} =   - \partial^{0}{} \omega^{mij} -  f^{0}{}^{j} \eta^{mi} + f^{0}{}^{i} \eta^{mj} + \partial^{m}\omega^{0}{}^{ij}=0
\end{align}
whose symmetric part is zero, and in the hook part we use $f^{0j}$, $\omega^{0ij}$ and $\omega^{mij}$ from (\ref{ef0m}),(\ref{ew0im}), (\ref{wabc}) respectively, and $\tau_m=0$ to obtain second constraint on the geometry
\begin{align}
&- \partial_{0}{} \partial^{i}\phi^{jm} + 2 \partial_{0}{} \partial^{j}\phi^{im}  -  \partial_{0}{} \partial^{m}\phi^{ij}=0
\end{align}
Contraction of the constraint with $h_{im}$ gives condition
\begin{align}
 \partial_{0}{} \partial_{i}\phi^{ji} =0
\end{align}
 which can be used later if necessary for simplifications. The antisymmetric part, after insertion of $\omega^{ijl}$, 
and the value for $\omega^{0}{}^{im}$ turns out to be exact equation.

The equation (\ref{et0j}) still has one gauge parameter to fix. One may define $\beta_m{}^m=0$ using $\xi^{0-}=-\partial_m\xi^{0,m}+h^0\xi^{m-}$. The symmetric and antisymmetric part of (\ref{et0j}) will identically vanish, while the hook part defines \begin{align}
    \kappa^n=\partial_m \beta^n{}^{m}.
\end{align}
The $m0$ component of the equation \begin{align}
E[B^i]^{m0}=\partial^m \beta^0{}^i-\partial^0\beta^m{}^i-h^{mi}\kappa^0+h_0{}^0f^{mi}=0    \label{eem0}
\end{align}
has antisymmetric part which makes relation $\partial^m\beta_0{}^i-\partial^i\beta_0{}^m=0$ symmetric, and symmetric part which allows us to define 
\begin{align}
     \kappa^{0}{}=-  \partial_{k}\partial^{l}\phi_{l}{}^{k} + \partial^{m}\partial_{m}\tau^{0}{}
\end{align}
the 0-th component of $\kappa_{\mu}$ field in terms of the $\tau^0$ and $\phi^{mn}$. 
It also allows to find $f^{mn}$ 
\begin{align}
    f^{mn}=   \partial_{0}{} \beta^{mn} + h^{mn} \partial_{k}\partial^{k}\tau^{0}{} - h^{mn} \partial_{l}\partial_{k}\phi^{kl} - \partial^{m}\partial^{n}\tau^{0}{}\label{efmn}
\end{align}
in terms of the fields $\phi^{mn}$, $\tau^0$ and $\beta^{mn}$. 
One can notice that all the fields can be determined in terms of the components of the $e_{\mu}{}^a$ and $\tau_{\mu}$. Where the remaining component of the $e_{\mu}{}^a$ is symmetric and traceless component while $\tau_{\mu}$ are the components that belong to H generator, i.e. both originated from relativistic $P_a$ generator of translations. The undetermined field $\beta^{mn}$ is symmetric and traceless field. 

The equations for the $mn$ and $m0$ components of the fields belonging to special conformal transformations generator $K^i$, (\ref{etim})
read {\footnotesize
\besubeqs
\begin{align}
     2 \partial_{0}{} \partial^{m}\beta^{ni} + \eta^{in} (\partial^{m}\partial_{a}\partial^{a}\tau^{0}{} - \partial^{m}\partial_{b}\partial_{a}\phi^{ab})  -  2\partial_{0}{} \partial^{n}\beta^{mi} -  \eta^{im} (\partial^{n}\partial_{a}\partial^{a}\tau^{0}{} + \partial^{n}\partial_{b}\partial_{a}\phi^{ab}) &=0,\label{eq145} \\
   - \partial_{0}\partial_0 \beta^{mi} + h^{im} \partial_{0}{} (- \partial_{a}\partial^{a}\tau^{0}{} + \partial_{b}\partial_{a}\phi^{ab}) + \partial_{0}{} \partial^{i}\partial^{m}\tau^{0}{} + \tfrac{1}{2} \partial^{i}\partial^{m}\partial_{a}e^{0}{}^{a}&=0
\end{align}\esubeqs } and equations for the $m,n$ and $m,0$ components of the generator for the special conformal transformations in the 0 direction (\ref{eti0}) are
\begin{align}
   \partial^{m}\partial_{e}\beta^{ne} -  \partial^{n}\partial_{e}\beta^{me}&=0 \\
- \partial_{0}{} \partial_{e}\beta^{me} + \partial^{m}\partial_{e}\partial^{e}\tau^{0}{} -  \partial^{m}\partial_{j}\partial_{e}\phi^{ej}&=0 \label{e143}
\end{align} respectively.
The latter can be combined with a (\ref{eq145}) into constraint on the field $\beta^{mn}$, $2\partial_0\partial^m\beta^{ni}+\eta^{in}\partial_0\partial_l\beta^{ml}-2\partial_0\partial^n\beta^{mi}-\eta^{im}\partial_0\partial_l\beta^{nl}=0$.
Taking the trace and the constraint $\beta^{m}{}_m=0$ one obtains that $\partial_0\partial_l\beta^{ml}=0$ vanishes. Inserting that into (\ref{e143}) gives relation between 
$\partial^m\partial^l\partial_l\tau^0=\partial_m\partial_j\partial_l\phi^{lj}$.
This, combined with $\delta_{im}$ contraction of the equation (\ref{eqap}) from the appendix  gives dynamical equation for the field $\phi^{mn}$
\begin{align}
    \partial_m\partial_i\partial^i\phi^{jn}-\partial^m\partial_i\partial^j\phi^{ni}-\partial^m\partial_i\partial^n\phi^{ji}=0.
\end{align}
We write the list of gauge fields in the Table (1) below, we also give a list of gauge transformations and gauge parameters in the appendix in Table (\ref{tablgcg})
\begin{center}
\begin{table}[ht!]
\begin{tabular}{|c|l|c|}
\hline
Generator &  Gauge field & Gauge parameter   \\ \hline
$D$ & $b^{m}=0$ & $\xi^{+-}$ \\
     & $b^{0}=\frac{1}{2}\partial_ie^{0i}$ &   \\ \hline
$P_i$  &  $e^{m}{}^i\equiv\phi^{mi}$  &  $\xi^{i+}$ \\
  &  $e^{0}{}^i$  &   \\ \hline
$H$  &  $\tau^{m}=0$  & $\xi^{0+}$ \\
  &  $\tau^{0}$  &  \\ \hline 
$J_{ij}$  & $\omega^{jm,n}=\partial^m\phi^{nj}-\partial^{n}\phi^{mj}$ &  $\xi^{i,j}$ \\ 
&  $\omega{}^{0mi} -  \partial^{0}{} \phi^{mi} -  b^{0}{} \eta^{mi} + \partial^{m}e^{0}{}^{i}=0$ &  \\  \hline
$B_i$ &  $\beta^{m}{}^i$ & $\xi^{0,i}$ \\
 &  $\beta^{0}{}^i=\partial^i\tau^0$ & \\ \hline
$K_i$ & $f^{mn}= \partial_{0}{} \beta^{mn} + h^{mn} \partial_{k}\partial^{k}\tau^{0}{} - h^{mn} \partial_{l}\partial_{k}\phi^{kl} - \partial^{m}\partial^{n}\tau^{0}{}$ & $\xi^{i-}$ \\  & $f^{0m}=\frac{1}{2}\partial^m\partial_i e^{0i}$  &  \\ \hline
$K_0$ & $\kappa^n=\partial_m\beta^{mn}$ & $\xi^{0-}$  \\  & $\kappa^0=-\partial_k\partial^l\phi_l{}^k+\partial^m\partial_m\tau^0$ &  \\\hline
\end{tabular}
\caption{ List of gauge fields with their corresponding generators and gauge parameters}
\end{table}
\end{center}
We have obtained four remaining fields $\phi^{mn}, e^{0i} ,\beta^{mn},\tau^0$, and two free gauge parameters $\xi^{m+},\xi^{0+}$  in terms of which the others are expressed. We also got one dynamical equation for the field $\phi^{mn}$, and constraints on the geometry which did not play a role in determination of the dynamical equation or expressing  the other gauge fields. Earlier analyses of the Carrollian theories showed that such constraints play a role of constraints for Lagrange multipliers in the form of the undetermined fields \cite{Bergshoeff:2016soe,Bergshoeff:2017btm}. 

\section{Spin-3}

We consider the irreducible Young module of $so(3,2)$ $\YoungpB$, which describes spin-3 field $t_{ab}=t_{ba}$. As in previous example of conformal graviton, to find the solution of the linearized equations and fix suitably gauge symmetries, we have to find the IW contraction analogous to the contraction above. The commutation relations of the generator of translations and the generators of spin three algebra are 
\begin{align}
    [P_a, t_{c+}]&=-\eta_{ac}t_{++}\,, & [P_a, t_{c-}]&=t_{ac}+\frac{1}{2}t\fdu{m}{m}\,, & [P_a, t_{--}]&=2t_{a-}\,, \\ [P_a, t_{++}]&=0\,, &
    [P_a, t_{cd}]&=-\eta_{ac}t_{d+}-\eta_{ad}\ttt_{c+}\,. \label{algp3}
\end{align}
For the conformal graviton case, it was necessary to choose $H,K_0$ and $B_i$ to be elements of $\mathfrak{i}$, so here we proceed in the same way, choosing the generators $SH,SK_0,SB_i$ to be elements of $\mathfrak{i}$ while the remaining generators build $\mathfrak{h}$
\begin{align}
    \mathfrak{h}&=\{ SP_j,SK_j,t_{++},t_{--},S_{jk},SC \}&&
    \mathfrak{i}=\{ SH,SK_0,SB_i \}.
\end{align}
The dictionary we use between the symmetric $t^{\Lambda}$ generators $S^{AB}$ for the spin-3 case, and the generators denoted here is 
$t^{j+}=SP^j,t^{j-}=SK^j,t^{jk}=SJ^{jk},t^{00}=SC$, also $t^{0+}=SH,t^{0-}=SK^0,t^{0i}=SB^i$. The letter $S$ in front of what one can recognize as generators in conformal Carrollian algebra, denote that they originate from the symmetric generators of $t^{\Lambda}$.  
The algebra that we obtain after taking the limit reads
\begin{align*}
    [P_i, SP_{j}]&=-\eta_{ij}t_{++}\,, & [P_i, SK_{j}]&=SJ_{ij}+\tfrac12 (SJ\fdu{k}{k}+SC)\eta_{ij}\,, & [P_i, t_{--}]&=2SK_{i-}\,, \\ [P_i, t_{++}]&=0\,, &
    [P_i, SJ_{jk}]&=-\eta_{ij}SP_{k}-\eta_{ik}SP_{j}\,, \\
[P_i, SH]&=0\,, & [P_i, SK_{0}]&=SB_{i}\,, & [P_i, SB_{k}]&=-\eta_{ik}SH\,, \\ [H, t_{++}]&=0\,, &
      [H, t_{--}]&=2SK_{0}\,, & [H, SK_{j}]&=B_{j}\,, \\
       [H, t_{--}]&=2SK_{0}\,,  &  [H, SJ_{jk}]&=0\,. & [H, SP_{j}]&=0\,
      \end{align*}
      so that  one can write the algebra valued one-form $\omega$ and zero-form $\xi$ as 
{\footnotesize      
\begin{align}
    \omega&= se^{i}SP_{i}+s\tau SH +\tfrac12 s\omega^{ij}SJ_{ij}+s\beta^{i}SB_{i}+\gamma SC+\frac12\omega^{++}t_{++}+\frac12\omega^{--}t_{--}+sf^{i}K_{i}+s\kappa K_{0}\,,\\
    \xi&= se^{i}\xi_{i+}+s\tau S\xi^{0+} +\tfrac12 s\omega_{ij}\xi^{ij}+s\beta_{i}\xi^{0i}+\gamma \xi^{00}+\frac12\omega^{++}\xi^{++}+\frac12\omega^{--}\xi^{--}+sf_{i}\xi^{i-}+s\kappa\xi^{0-}\,.
\end{align} }      
The system of linearized equations and gauge symmetries reads
\besubeqs
\begin{align}
    t_{++}&: &d\omega^{++}-2h_k\wedge se^{k}&=0\,, &\delta\omega^{++}&= d\xi^{++}-2h_k\xi^{k+}\,,\label{stpp} \\
    SP_{i}&: &d se^{i}-h_l \wedge \omega^{li}&=0\,,
    &\delta se^{i}&=d\xi^{i+}-h_l \xi^{li}\,, \label{stip}\\
    SH&: &ds\tau-h_i \wedge s\beta^{i}-h^0\wedge\gamma&=0\,, &\delta s\tau&=d\xi^{0+}-h_l \xi^{l0}-h^0\xi^{00}\,,\label{st0p}\\
    SJ_{ij}&: &d\omega^{ij}+h^{(i} \wedge sf^{j)}+h_l\wedge sf^{l}\eta^{ij}&=0\,, &\delta \omega^{ij}&=d\xi^{ij}+h^{(i} \xi^{j)-}+h_l\xi^{l-}\eta^{ij}\,,\label{stij}\\ 
SB_{j}&: &ds\beta^{j}+h^{0} \wedge sf^{j}+h^{j} \wedge s\kappa&=0\,, &\delta s\beta^{j}&=d\xi^{0j}+h^{0} \xi^{j-}+h^{j} \xi^{0-}\,,\label{st0j}\\ 
SC&: & d\gamma+h^i\wedge sf_i&=0\,, & \delta\gamma&=d\xi^{00}+h^i\xi_i{}^-\,, \label{st00}\\
    SK_{i}&: &dsf^i+h^i\wedge\omega^{--}&= 0\,, &\delta sf^{i}&=d\xi^{i-}+h^i\xi^{--}\,,  \label{stim}\\
SK_{0}&: &ds\kappa+h^0\wedge\omega^{--}&= 0\,, &\delta s\kappa&=d\xi^{0-}+h^0\xi^{--}\,, \label{st0m}\\
     st_{--}&: &d\omega^{--}&=0\,, & \delta\omega^{--}&=d\xi^{--}\,.\label{stmm}
\end{align}
\esubeqs
From the first gauge condition
(\ref{stpp})  we see that we can fix  $\omega^m{}^{++}=0$ to be equal to zero,  by setting $\xi^m{}^+=\frac{1}{2}\partial^m\xi^{++}$. Since the component of the background field $h_{k0}$ vanishes, the field $\omega^0{}^{++}$ remains unfixed, and its gauge transformation is set to $\delta\omega^0{}^{++}=\partial^0\xi^{++}$. 
The m and n component of the equation require from the field $se^{mn}=se^{nm}$ to be symmetric, and $m0$ component defines \begin{align}
    se^{0m}=\frac{1}{2}\partial^m\omega^{0++}
\end{align} 

From the equation and gauge condition
(\ref{stip}) we can determine that the remaining symmetric part of the field $se^{(mi)}=0$ is zero, similarly as in non-relativistic case, by fixing $\xi^{mi}=\partial^m\xi^{i+}=\frac{1}{2}\partial^m\partial^i\xi^{++}$. The zero-th component gauge condition fixes gauge transformation of $\delta se^{0i}=\partial^0\xi^{i+}=\frac{1}{2}\partial_0\partial^i\xi^{++}$.
The symmetric and antisymmetric $mn$ component of the equation are zero, and only the hook component gives the condition $\omega^{mni}=\omega^{nmi}$ which makes $\omega^{mni}\equiv\phi^{mni}$ totally symmetric. The $m0$ component's symmetric and antisymmetric part fix $ \omega^{0mi}$ field in terms of the $\partial^i se^{0m}$
\begin{align}
    \omega^{0mi}=\frac{1}{2}(\partial^m se^{0i}+\partial^i se^{0m}), && \partial^m se^{0i}=\partial^i se^{0m}.
\end{align}
 and yield this to be a symmetric tensor.

From the gauge condition 
(\ref{st0p}) we can already see the second gauge parameter which will not fix any field components $\xi^{0+}$. The remaining two gauge parameters, $\xi^{0m}=\partial^m\xi^{0+}$ and $\xi^{00}=\partial^0\xi^{0+}$ fix $s\tau^m=0$ and $s\tau^0=0$ respectively. The $mn$ component of the equation define $s\beta^{mn}=s\beta^{nm}$ to be symmetric field and $s\beta^{0m}=h^{00}\gamma^m$ the zero-th component to be equal to field $\gamma^m$.
The gauge condition and the equation from the $SJ_{ij}$ generator (\ref{stij}) determine the gauge parameter $\xi^{i-}=-\frac{2}{5}\partial_m\xi^{im}$ 
and the trace of the field  $\phi_m{}^{im}=0$. 
One can write the gauge transformation for the two dimensional field $\phi^{mni}$ as
 \begin{align}
     \delta\omega^{mij}=\frac{1}{2}\partial^m\partial^j\partial^i\xi^{++}-\frac{1}{10}(\partial^l\partial^j\partial_l\xi^{++}h_m{}^i+\partial^l\partial^i\partial_l\xi^{++}h_m{}^j)-\frac{1}{5}\partial_l\partial^m\partial^l\xi^{++}\eta^{ij}.
 \end{align} 
Due to fixing the trace of $sf_i{}^i=0$ with $\xi^{--}=-\frac{1}{2}\partial_i\xi^{i-}$, the only non-vanishing component, hook component of the equation gives $\partial_m\omega_i{}^{im}=0$, and
\begin{align}
     sf^{nm} = -   \partial_{i}\phi^{mni}. \end{align}
Without taking contractions, equation (\ref{stij}) gives equation for the $\phi^{mni}$ field \begin{align}\tfrac{1}{6} (- \eta^{mn} \partial_{a}\phi^{ija} + \eta^{jn} \partial_{a}\phi^{ima} + \eta^{jm} \partial_{a}\phi^{ina} -  \eta^{in} \partial_{a}\phi^{jma} -  \eta^{im} \partial_{a}\phi^{jna} \nonumber \\ +  \eta^{ij} \partial_{a}\phi^{mna} + \partial^{i}\phi^{jmn} - 3 \partial^{j}\phi^{imn} + \partial^{m}\phi^{ijn} + \partial^{n}\phi^{ijm})=0.\label{s3cond1}\end{align}
The $m0$ component of the equation have only hook and symmetric part of the equation. They can be solved to determine 
\begin{align}
    sf^{0j}=-\frac{2}{5}\partial_i\omega^{0ji}=-\frac{1}{5}\partial_i\partial^i\partial^j\omega^{0++}.
\end{align}
From (\ref{st0j}) gauge condition we  fix $s\beta^i{}_i=0$ with $ \xi^{0-}$, we have thus 
\begin{align}s\kappa^m=\partial_{j} s\beta^{mj} .\end{align} The $m0$ component gives $\delta s\beta^{0j}=\partial_0\xi^{0j} +\xi^{j-}$ and $s\kappa^0=\frac{1}{2}\partial_j s\beta^{0j}$.
Components $mn$ and $m0$ from (\ref{stim})  give
\begin{align}
    \omega^{n--} =- \partial_{i}sf^{ni}=\partial_i\partial_j\phi^{nij}  &&  \omega^{0--}{} =-\frac{1}{2} \partial_{i}sf^{0i}=\frac{1}{10}\partial_i\partial^i\partial_j\partial^j\omega^{0++}.
\end{align} 
The $mn$ component of (\ref{st0m}) is exactly satisfied, while $m0$ component gives constraint on the $ s\beta^{ij}$ and $\phi^{ijk}$ fields $-\frac{1}{2}\partial^m\partial_j s\beta^{0j}+\partial_0\partial_ls\beta^{ml}-\partial_i\partial_j\phi^{imj}=0$.
The equation (\ref{stmm}) gives condition on the $\phi^{mni}$ field $\epsilon_{mn}\partial_i\partial_j\phi^{inj}=0$ and condition on $\omega_0{}^{++}$, $\frac{1}{10}\partial^m\partial_i\partial^i\partial_l\partial^l\omega_0{}^{++}-\partial_0\partial_i\partial_j\phi^{mij}=0$. Using the constraint from (\ref{stmm}) and from (\ref{st0m}) one obtains $\epsilon_{mn} \partial^m\partial_j s\beta^{0j}=0$.

To present the results transparently, we write the gauge fields with corresponding gauge parameters and generators in the Table (\ref{tab2}) below, while the list of linearized gauge transformations and expressions for gauge parameters, we give in the appendix, in the Table (\ref{tab4}).
\begin{center}
\begin{table}[ht!]
\begin{center}
\begin{tabular}{|c|l|c|}
\hline
Generator &  Gauge field & Gauge parameter   \\ \hline
$St_{++}$ & $\omega_m{}^{++}=0$ & $\xi^{++}$ \\
     & $\omega_{0}^{++}$ &    \\ \hline
$SP_i$  &  $se_{mn}=0$  &  $\xi^{i+}$ \\
  &  $se^{0m}=\frac{1}{2}\partial_m\omega_0{}^{++}$  &   \\ \hline
$SH$  &  $s\tau^{m}=0$  & $\xi^{0+}$ \\
  &  $s\tau^{0}=0$  &  \\ \hline 
$SJ_{ij}$  & $\omega^{mni}=\phi^{mni}$ &  $\xi^{ij}$ \\ 
&  $\omega{}^{0mi}=\frac{1}{2}\partial^m\partial^i\omega_0{}^{++}$&  \\  \hline
$SB_i$ &  $ s\beta^{m}{}^i$ & $\xi^{0i}$ \\
 &  $ s\beta^{0}{}^i=\gamma_m$ & \\ \hline
$SK_i$ & $sf^{mn}= -\partial_{i}\phi^{mni}$ & $\xi^{i-}$ \\  & $sf^{0m}=-\frac{1}{5}\partial_i\partial^i\partial^m\omega_0{}^{++}$  &  \\ \hline
$SK_0$ & $s\kappa^n=-\partial_m s\beta^{mn}$ & $\xi^{0-}$  \\  & $s\kappa^0=-\frac{1}{2}\partial_j s\beta^{0j}$ &
\\ \hline $St_{--}$ & $\omega_n{}^{--}=\partial_i\partial_j\phi^{nij}$&  $\xi^{--}$\\  & $\omega_0{}^{--}=\frac{1}{10}\partial_i\partial^i\partial_j\partial^j\omega_0{}^{++}$ & 
 \\\hline
 $SC$ & $\gamma_m$ & $\xi^{00}$ \\  & $\gamma_0$ & 
\\\hline
\end{tabular}
\end{center}
\caption{List of gauge fields with corresponding generators and gauge parameters for the spin-3 conformal gravity in ultra-relativistic limit}\label{tab2}
\end{table}
\end{center}
Similarly like in the previous case, we have obtained one dynamical field $\phi^{mni}$ with corresponding dynamical equation and four additional remaining fields $\omega^{0++}, s\beta^{mi}$, $\gamma_m$ and $\gamma_0$. We also obtained constraints on the geometry, which should play similar role as in the spin-2 case. Analogously one may expect that some of them correspond to radiative degrees of freedom for gravity at null-infinity.

\textbf{Comment on generalization to spin-s.}

To determine the spectrum of the theory for the arbitrary spin, one first has to perform IW contraction of the given algebra. For the spin-s, the algebra prior to contraction is going to be defined by the Young tableaux with the s-1 boxes in the first row and s-t boxes in the second row. The generators that need to be set in the $\mathfrak{h}$, for analogous consideration as in spin-2 and spin-3 cases here, are those that have 1 and 2 components, and those that appear in the definition of traces. That way one makes sure that the algebra $\mathfrak{h}$ is closed. The generators in the 0-th direction have to be part of $\mathfrak{i}$. After solving the system of equations,
one can expect to obtain the lower dimensional dynamical spin-s field, and minimally two remaining fields which are undetermined, but are related with constraints on the geometry, and play a role of radiative degrees of freedom at null-infinity. One should also expect to obtain two remaining gauge parameters, that determine linear gauge transformations of the remaining fields.

\section{Conformal Carroll Spin-2 and Spin-3 Gravity}

 Conformal Carrol gravity is defined by the Chern-Simons action (\ref{csac3d}) and the connection (\ref{connection}). 
 We want to see what kind of generalizations one obtains from conformal Carroll algebra in comparison to the Carroll algebra.
 Carroll gravity for the flat background has a line element \begin{align}
     ds^2=\rho^2d\phi^2+d\rho^2
 \end{align}
defined by $e^a e^b\delta_{ab}$ where the corresponding zweibeins take the form $e_{\phi}^1=\rho, e_{\rho}^2=1$ and $e_{\rho}^2=e_{\phi}^2=0$.
Here, $\rho$ denotes radial and $\phi$ angular coordinate periodic in $2\pi$, $\phi\sim\phi+2\pi$. The time component is $\tau=dt$. 
Following the known procedure \cite{Campoleoni:2011hg,Banados:1998ta,Grumiller:2016pqb} we partially fix the radial gauge and consider the connection in the form
\begin{align}
    \omega=b^{-1}(\rho)(d+\mathcal{\omega}(t,\phi))b(\rho).\label{cong}
\end{align}
We allow for the connection on the boundary to depend on $t$ and $\phi$ coordinates, but not on $\rho$ coordinate. It is convenient to impose the following boundary conditions 
{\footnotesize
\besubeqs
\begin{align}
    \omega_{\phi}^{(s_2)}&=B_i+\mathcal{H}(t,\phi)H+\mathcal{P}_i(t,\phi)P_i+\mathcal{J}_{ij}(t,\phi)J_{ij}+\mathcal{K}_i(t,\phi)K_i+\mathcal{K}_0(t,\phi)K_0+\mathcal{D}(t,\phi)D \label{bcs2f}\\
    \omega_t^{(s_2)}&=\mu(t,\phi)H \label{bcs2t}
\end{align}
\esubeqs}
denoting with $(s_2)$ that we are considering spin-2 case.
We allow for the functions from $\omega_{\phi}^{(s_2)}$ to vary at the boundary, while variation of the chemical potential $\delta\mu(t,\phi)=0$ is fixed. 

To find the metric formulation for these boundary conditions we define the group element which needs to be valued in the algebra, $b(\rho)=e^{\rho P_2}$. 
When the generator in the group element commutes with generators from the boundary conditions into generator of translations, one obtains contribution to zweibein.
Using the Baker-Cambell-Hausdorff formula, we can determine connection (\ref{cong}) and read out the components of the zweibein and $\tau$ to be
\besubeqs
\begin{align}
    e_{\phi}^1&=\mathcal{J}_{12}(t,\phi)\rho+\mathcal{P}_1(t,\phi)\,, && e_{\phi}^2=\mathcal{P}_2(t,\phi)-\mathcal{D}(t,\phi)\rho \\
    e_{\rho}^1&=0\,, && e_{\rho}^2=1 \\
    \tau_t&= \mu(t,\phi)\,, && \tau_{\phi}=\rho+\mathcal{H}(t,\phi)
\end{align}
\esubeqs
That leads to the line element {\footnotesize \begin{align}
    ds^2=\left((\mathcal{P}_2(t,\phi) + \mathcal{D}(t,\phi)\rho)^2 + (\mathcal{P}_1(t,\phi) + \mathcal{J}_{12}(t,\phi) \rho)^2\right)d\phi^2+2(\mathcal{P}_2(t,\phi) + \mathcal{D}(t,\phi) \rho)d\phi d\rho+d\rho^2\label{eq55} \end{align} }
and 
\begin{align}
    d\tau=(\rho+\mathcal{H}(t,\phi))d\phi+\mu(t,\phi)dt\label{dtccg}
\end{align}
which allows for the two additional functions to appear in the two dimensional line element in conformal Carroll gravity (\ref{eq55}), compared to similar line element in Carroll gravity \cite{Bergshoeff:2016soe}. 
In general, conformal Carroll gravity is expected to allow for the additional functions in the line element compared to Carroll gravity. However, even with these additional functions, one cannot eliminate the shift in the $\phi$ direction, the same as in Carroll gravity. 

We wonder which boundary conditions could be imposed in the spin-3 case such that we do not have shift in the $\phi$ component in the (\ref{dtccg}).
Following the same procedure as above, we partially fix to radial gauge and write the connection in the form (\ref{cong}) which can depend on $t$ and $\phi$ on the boundary, but not on coordinate $\rho$. The connection $\omega(t,\phi)$ is now of the form $\omega(t,\phi)=\omega(t,\phi)^{(s_2)}+\omega(t,\phi)^{(s_3)}$. Where $(s_2)$ represents boundary conditions for conformal gravity (spin-2 case), i.e. equations (\ref{bcs2f}) and (\ref{bcs2t}). The additional minimal boundary condition that we impose due to spin-3 is 
\begin{align}
    \omega_{\phi}^{(s_3)}=\mathcal{S}(t,\phi)t_{++}\label{eq57}
\end{align}
and the group element $b(\rho)$ that we choose is $b(\rho)=e^{\rho P_2+a_2\rho SP_2+a_3B_1}$. The components of the metric for such extension of conformal Carroll gravity will have components
\besubeqs
\begin{align}
    e_{\phi}{}^1&=(\mathcal{P}_1(t,\phi)+\mathcal{J}_{12}(t,\phi)-a_3\mathcal{P}_1(t,\phi)) && e_{\phi}=\mathcal{P}_2(t,\phi)+\mathcal{D}(t,\phi)\rho \\
    e_{\rho}{}^1&=0  &&  e_{\rho}^2=1 \\
    \tau_{\phi}&=\mathcal{H}(t,\phi)+\rho+2a_2\mathcal{S}(t,\phi)\rho+a_3\mathcal{P}_1(t,\phi) && \tau_t=\mu(t,\phi)
\end{align}
\esubeqs
from which we see that suitable condition on $\mathcal{P}_1$ and $\mathcal{S}$ will lead to disappearance of the shift in the $\phi$ direction. For that we had to not only add (\ref{eq57}) minimal boundary condition, but also modify the group element $b(\rho)$.

\section{Conclusion}
We have studied IW contraction of the algebras corresponding to the Young modules $\YoungpAA$ and $\YoungpB$ from the $so(3,2)$ decomposition of $hs(\YoungpA)$. First of which described conformal graviton and the second spin-3 field. The choice of IW contraction that we took, led in the first case to the conformal Carrollian algebra, ultra-relativistic limit of the conformal algebra. Based on this algebra we have constructed an ultra-relativistic gravity theory starting from the Chern-Simons action. For the spin-3 case, we have conducted IW contraction along the same lines as for the conformal spin-2 graviton, followed by the construction of the gravity theory based on Chern-Simons action. It would be conceivable to think of it as a possible spin-3 generalization of the conformal Carroll gravity.

We have also studied holography of the conformal Carroll gravity imposing the suitable set of boundary conditions. 
In the metric formulation this boundary conditions led to the generalized line element compared to the one obtained from Carroll gravity. 
The line element contained the $d\tau$ part with a shift in the $\phi$ direction which was not possible to  remove within Carroll or conformal Carroll gravity. Considering the holography of spin-3 theory, we managed to find suitable projector and boundary conditions which eliminate this shift. 

As a future direction of study it would be interesting to consider the spin-s generalization of the map between the conformal Carroll algebras and the higher spin fields in the metric formulation. It would be especially interesting to see if formulation of three dimensional conformal higher spin gravity in this limit would give additional information on the restriction of interacting vertices in higher spin theory.

\section*{Acknowledgments}
\label{sec:Aknowledgements}

The author would like to thank Jan Rosseel, Daniel Grumiller, Romain Ruzziconi and Adrien Fiorucci for the discussions.
This work was made possible by the Hertha Firnberg grant T 1269-N by Austrian Science fund (FWF).
\section{Appendix}

\subsection{Conformal gravity example}
Inserting the $f^{im}$ (\ref{efmn}) into the first constraint from the equation (\ref{fc}) one obtains 
\begin{align}
   & \tfrac{1}{2} (-2 \partial_{0}{} \beta^{jn} \eta^{im} + 2 \partial_{0}{} \beta^{jm} \eta^{in} + 2 \partial_{0}{} \beta^{in} \eta^{jm} - 2 \partial_{0}{} \beta^{im} \eta^{jn} + 4 \eta^{in} \eta^{jm} \partial_{a}\partial^{a}\tau^{0}{} \nonumber \\ & - 4 \eta^{im} \eta^{jn} \partial_{a}\partial^{a}\tau^{0}{} - 4 \eta^{in} \eta^{jm} \partial_{b}\partial_{a}\phi^{ab} + 4 \eta^{im} \eta^{jn} \partial_{b}\partial_{a}\phi^{ab} + \partial^{i}\partial^{m}\phi^{jn} + 2 \eta^{jn} \partial^{i}\partial^{m}\tau^{0}{} \nonumber \\ &-  \partial^{i}\partial^{n}\phi^{jm} - 2 \eta^{jm} \partial^{i}\partial^{n}\tau^{0}{} -  \partial^{j}\partial^{m}\phi^{in} - 2 \eta^{in} \partial^{j}\partial^{m}\tau^{0}{}  + \partial^{j}\partial^{n}\phi^{im} + 2 \eta^{im} \partial^{j}\partial^{n}\tau^{0}{} \nonumber \\ &+ \partial^{m}\partial^{i}\phi^{jn} -  \partial^{m}\partial^{j}\phi^{in} -  \partial^{n}\partial^{i}\phi^{jm} + \partial^{n}\partial^{j}\phi^{im})=0\label{eqap}
\end{align}
\subsection{Gauge parameters and gauge transformations}
In this Table \ref{tablgcg} we present the gauge parameters and linearized gauge transformations for the spin-2 case
\begin{center}
\begin{table}[ht!]
\begin{tabular}{|c|l|l|c|}
\hline
Generator &  Gauge & linearized gauge transformation & Gauge parameter  \\ & field & &   \\ \hline
$D$ & $b^{m}$ & $ \delta b^m=0$ & $\xi^{+-}=\frac{1}{2}\partial_l\xi^{l+}$ \\
     & $b^{0}$ & $  \delta b_0=\frac{1}{2}\partial_0\partial_l\xi^{l+} $ & \\ \hline
$P_i$  &  $e^{m}{}^i$ & $ \delta e^{mi}=\frac{1}{2}(\partial^m\xi^{i+}+\partial^i\xi^{m+})-\frac{1}{2}h^{mi}\partial_l\xi^{l+}$  &  $\xi^{i+}$ \\
  &  $e^{0}{}^i$ & $ \delta e^{0i}=\partial^0\xi^{i+}$ &   \\ \hline
$H$  &  $\tau^{m}$ & $\delta\tau_m=0$ & $\xi^{0+}$ \\
  &  $\tau^{0}$  &  $\delta\tau^{0}=\partial^0\xi^{0+}-\frac{1}{2}h_0{}^0\partial_l\xi^{l+}$  & \\ \hline 
$J_{ij}$  & $\omega^{jm,n}$ & $ \delta\omega^{m i,j}=\frac{1}{2}\partial^m(\partial^j\xi^{i+}-\partial^i\xi^{j+})$ &  $\xi^{i,j}=\frac{1}{2}(\partial^{j}\xi^{i+}-\partial^{i}\xi^{j+})$ \\ 
& & \hspace{1.5cm}$-\frac{1}{2}\partial_l(h^{mi}\partial^j\xi^{l+}-h^{jm}\partial^i\xi^{l+}) $& 
\\
&  $\omega{}^{0mi} $ & $ \delta\omega^{0i,j}=\frac{1}{2}\partial^0(\partial^j\xi^{i+}-\partial^{i}\xi^{j+})$ &  \\  \hline
$B_i$ &  $\beta^{m}{}^i$ & $\delta\beta^{mj}=-\partial^m\partial^j\xi^{0+}+\frac{1}{2}h^{jm}\partial_l\partial^l\xi^{0+}$ & $\xi^{0,i}=\partial^i\xi^{0+}$ \\
 &  $\beta^{0}{}^i$ & $\delta\beta^{0j}=-\partial^0\partial^j\xi^{0+}+\frac{1}{2}h_0{}^0\partial^j\partial_l\xi^{l+}$ & \\ \hline
$K_i$ & $f^{mn}$ & $\delta f^{mi}=\frac{1}{2}\partial^m\partial^i\partial_l\xi^{l+}$ & $\xi^{i-}=-\frac{1}{2}\partial^i\partial_l\xi^{l+}$ \\  & $f^{0m}$ & $\delta f^{0i}=\frac{1}{2}\partial^0\partial^i\partial_l\xi^{l+}$  &  \\ \hline
$K_0$ & $\kappa^n$&  $\delta\kappa_m=-\frac{1}{2}\partial_m\partial_l\partial^l\xi^{0+}$ & $ \xi^{0-}=-\frac{1}{2}\partial_m\partial^m\xi^{0+}$  \\  & $\kappa^0$ & $\delta\kappa^0=-\frac{1}{2}\partial^0\partial_l\partial^l\xi^{0+}$ & \\\hline
\end{tabular}
\caption{List of linearized gauge transformations of the fields and gauge parameters} \label{tablgcg}
\end{table}
\end{center}
and in the Table \ref{tab4} for the spin-3 case
\begin{center}
\begin{table}[ht!]
\begin{tabular}{|c|l|l|c|}
\hline
Generator &  Gauge  &Linearized gauge transformations & Gauge parameter\\ & field& &  \\ \hline
$St_{++}$ & $\omega_m{}^{++}$ &  $ \delta\omega^m{}^{++}=0$ & $\xi^{++}$ \\
     & $\omega_{0}^{++}$ & $\delta\omega_0^{++}=\partial_0\xi^{++}$ &  \\ \hline
$SP_i$  &  $se_{mn}$  & $\delta se_{mn}=0$& $\xi^{m+}=\frac{1}{2}\partial^m\xi^{++}$\\
  &  $se^{0m}$  & $\delta se^{0i}=\frac{1}{2}\partial^0\partial^i\xi^{++}$ & \\ \hline
$SH$  &  $s\tau^{m}$ & $\delta s\tau_m=0$ & $\xi^{0+}$ \\
  &  $s\tau^{0}$  & $\delta s\tau_0=0$ & \\ \hline 
$SJ_{ij}$  & $\omega^{mni}$ & $ \delta\omega^{mij}=\frac{1}{2}\partial^m\partial^j\partial^i\xi^{++}+C\partial_l\partial^m\partial^l\xi^{++}\eta^{ij} $ & $\xi^{mi}=\frac{1}{2}\partial^m\partial^i\xi^{++}$  \\ 
&  &\hspace{1.5cm}$+\frac{C}{2}(\partial^l\partial^j\partial_l\xi^{++}h_m{}^i+\partial^l\partial^i\partial_l\xi^{++}h_m{}^j)$ &\\
&  $\omega{}^{0mi}$& $\delta\omega{}^{0mi}=\frac{1}{2}\partial^0\partial^i\partial^m\xi^{++}$ &  \\  \hline
$SB_i$ &  $\beta^{m}{}^i$ & $\delta\beta^{m}{}^i=\partial^m\partial^i\xi^{0+}-\frac{1}{2}h^{mi}\partial^l\partial_l\xi^{++}$ &  $\xi^{0m}=\partial^m\xi^{0+}$ \\
 &  $\beta^{0}{}^i$ &$\delta\beta^{0}{}^i=\partial^0\partial^i\xi^{0+}+Ch_0{}^0\partial^i\partial^l\partial_l\xi^{++}$ & \\ \hline
$SK_i$ & $sf^{mn}$ & $\delta sf^{mi}=C\partial^m\partial_l\partial^l\partial^i\xi^{++}-\frac{C}{2}h^{mi}\partial_j\partial_n\partial^n\partial^j\xi^{++}$ & $\xi^{i-}=C\partial_m\partial^m\partial^i\xi^{++}$ \\  & $sf^{0i}$  & $\delta sf^{0i}=C\partial^0\partial_l\partial^l\partial^i\xi^{++}$ & \\ \hline
$SK_0$ & $s\kappa^n$ & $\delta s\kappa^n=-\frac{1}{2}\partial^n\partial_l\partial^l\xi^{0+}$ & $\xi^{0-}=-\frac{1}{2}\partial^m\partial_m\xi^{0+}$  \\  & $s\kappa^0$ &  $\delta s\kappa^0=-\frac{1}{2}\partial^0\partial_l\partial^l\xi^{0+}-\frac{C}{2}h_0^0\partial_i\partial_m\partial^m\partial^i\xi^{++}$ &
\\ \hline $St_{--}$ & $\omega_n{}^{--}$& $\delta\omega^{n--}=-\frac{C}{2}\partial^n\partial_i\partial_m\partial^m\partial^i\xi^{++}$ & $\xi^{--}=-\frac{C}{2}\partial_i\partial_m\partial^m\partial^i\xi^{++}$\\  & $\omega_0{}^{--}$ & $\delta\omega^{0--}=\frac{1}{10}\partial^0\partial_i\partial_m\partial^m\partial^i\xi^{++}$ &
\\ \hline
$St^{00}$ & $\gamma_m $ & $ \delta\gamma_m=\partial_m\partial^0\xi^{0+}+C\partial_l\partial^l\partial_m\xi^{++}$ & $h_0{}^0\xi^{00}=\partial_0\xi^{0+}$ \\
&$\gamma^0$& $\delta\gamma^0=\partial^0\partial^0\xi^{0+}$ &
 \\ \hline
\end{tabular}\caption{List of linearized gauge transformations and gauge parameters for the ultra-relativistic limit of the spin-3 conformal gravity, $C=-\frac{1}{5}$}\label{tab4}
\end{table}
\end{center}
\newpage
\footnotesize
\bibliographystyle{templatebib}
\bibliography{megabib.bib}

\end{document}